# Chapter 1
# Sonic Interactions in Virtual Environments: the Egocentric Audio Perspective of the Digital Twin

Michele Geronazzo and Stefania Serafin

**Abstract** The relationships between the listener, physical world and virtual environment (VE) should not only inspire the design of natural multimodal interfaces but should be discovered to make sense of the mediating action of VR technologies. This chapter aims to transform an archipelago of studies related to sonic interactions in virtual environments (SIVE) into a research field equipped with a first theoretical framework with an inclusive vision of the challenges to come: the egocentric perspective of the auditory digital twin. In a VE with immersive audio technologies implemented, the role of VR simulations must be enacted by a participatory exploration of sense-making in a network of human and non-human agents, called actors. The guardian of such locus of agency is the auditory digital twin that fosters intra-actions between humans and technology, dynamically and fluidly redefining all those configurations that are crucial for an immersive and coherent experience. The idea of entanglement theory is here mainly declined in an egocentric-spatial perspective related to emerging knowledge of the listener's perceptual capabilities. This is an actively transformative relation with the digital twin potentials to create movement, transparency, and provocative activities in VEs. The chapter will contain an original theoretical perspective complemented by several bibliographical references and links to the other book chapters that have contributed significantly to the proposal presented here.

Michele Geronazzo
University of Udine, Dept. of Humanities and Cultural Heritage, Udine, Italy
Imperial College London, Dyson School of Design Engineering, London, United Kingdom
e-mail: m.geronazzo@imperial.ac.uk

Stefania Serafin
Aalborg University Copenhagen, Dept. of Architecture, Design, and Media Technology, Copenhagen, Denmark, e-mail: sts@create.aau.dk

## 1.1 Introduction

Our daily auditory experience is characterized by immersion from the very beginning of our life inside the womb, actively listening to sounds surrounding us from different positions in space. Auditory information takes the form of a binaural continuous stream of messages to the left and right ears, conveying a compact representation of the omnidirectional source of learning for our existence [19, 48]. Both temporal and spatial activity of sounds of interest (e.g. dialogues, alarms, etc.) allow us to localize and encode the contextual information and intentions of our social interaction [1].

The hypothesis that our daily listening experience of sounding objects with certain physical characteristics dynamically shapes the acoustic features for which we ascribe meaning to our auditory world is supported by one of the key concepts in Husserl's phenomenology *"Meaning-bestowal"* ("Sinngebung" in German [73]) and by studies in ecological acoustics such as [48, 54, 96]. In particular, the idea of acoustic invariant as a complex pattern of change for a real-world sound interaction is strongly related to human perceptual learning and a socio-cultural mediation dictated by the real world. For some surveys of classical studies on the topic of ecological acoustics refer to [112].

From this perspective, acoustic invariants are learned on an individual basis through experiential learning. Hence, there is the need to trace their development over multiple experiences and to formalize a common ground for a dynamic expansion of individual knowledge. Any emerging understanding should be transferred to a technological system able to provide an immersive and interactive simulation of a sonic virtual environment (VE). Such a process must be adaptive and dynamic to ensure a level of coupling between user and system in such a way that the active listening experience is considered authentic.

Immersive virtual reality (here we generically referred to as VR) technologies allow immense flexibility and increasing possibilities for the creation of VEs with relationships or interactions that might be ontologically relevant even if radically different from the physical world. This can be evident by referring to the distinction between naturalistic and magical interactions, where the latter can be considered observable system configurations in the domain of artificial illusions, incredibly expanding the spectrum of possible digital experiences [13, 127].

One of the main research topics in the VR and multimedia communities is rendering. For decades, computer-aided design applications have favored -in the first place- the development of computer graphics algorithms. Some of these approaches, e.g. geometric ray-tracing methods, have been adapted to model sound propagation in complex VEs (see Ch. 3 for more details). However, there has been a clear tendency to prioritize resources and research on the visual side of virtual reality, confining auditory information to a secondary and ancillary role [158]. Although sound is an essential component of the grammar of digital immersion, relatively little compared to the visual side of things has been done to investigate the role of auditory space and environments. Nowadays, there is increasing consensus towards the essential contribution of spatial sound, also in (VR) simulations [9, 102, 145]. Technologies for **spatial audio rendering** are now able to convey perceptually plau-sible simulations with stimuli that are reconstructed from real-life recordings [18] or historical archives, as for the Cathédrale Notre-Dame de Paris before and after the 2019 fire [79], getting closer to a virtual version indistinguishable from the natural reality [77]. This is made possible by a high level of personalization in modeling user morphology and acoustic transformations caused by the human body interacting with the sound field generated in room acoustic computer simulations [17, 78, 114].

Nowadays, the boundary between technology and humans has increasingly blurred thanks to recent developments in research areas such as virtual and augmented reality, artificial intelligence, cyber-physical systems, and neuro-implants. It is not possible to easily distinguish where the human ends and the technology begins. For this reason, we embrace the idea of [10] who sees technology as a lens for the understanding of what it means to be human in a changing world. We can therefore consider the *phenomenal transparency* [94] where technology takes on the role of a transparent mediator for self-knowledge. According to Loomis [88], the **phenomenology of presence** between physical and virtual environments places the internal listener representation created by the spatial senses and the brain on the same level. Human-technology-reality relations are thus created by enactivity that allows a fluid and dynamic entanglement of all the involved actors.

In this chapter, we initially adopt Slater's definition of presence for an immersive VR system [135] embracing the recent revision by Skarbez [134]. The concepts of plausibility illusion and place illusion are central to capturing the subjective internal states. While the plausibility illusion determines the overall credibility of a VE in terms of subjective expectations, the place illusion establishes the quality of having sensations of being in a real place. They are both fundamental in providing credibility to a digital simulation based on individual experience and expectations concerning an internal frame of reference for scenes, environments, and events.[1]

We propose a theoretical framework for the new field of study Sonic Interactions in Virtual Environments (SIVE). We suggest from now on a unified reading of this chapter with references and integrations from all chapters of the corresponding book [49]. Each chapter provides state of the art, challenges and case studies for specific SIVE-related topics curated by internationally renowned scientists and their collaborators. The provided point of view focuses on the relations between real auditory experience and technologically mediated experiences in immersive VR. The first is characterized by individuality to confer immersiveness within a physical world. It is important to emphasize the omnidirectionally of auditory information that allows the listener to collect both the whole and the parts at 360°. The individualized auditory signals are the result of the acoustic transformations made by the head, ear, and torso of the listener that act as a spatial fingerprint for a complex spatio-temporal signal. Familiarity, and therefore previous experience with sounds, shape spatial localization capabilities with high intersubjectivity. Finally, studies on neural plasticity of the human brain confirm continuous adaptability of listening with impaired physiological functions, e.g. a hearing loss, and with electrical stimulation, e.g. via cochlear implants [82].

[1] For a dedicated discussion on the basic notions related to presence, please refer also to Ch. 11 in this volume.

The **mediated VR experience** is often characterized by the user's digital counterpart called avatar. It allows the creation of an embodied and situated experience in digital VEs. The scientific literature supports the idea that the manipulation of VR simulations can induce changes at the cognitive level [124], such as in educational [34] and therapeutic [106] positive effects. The ability of VR technologies to mediate within the immersive environment in embodied and situated relations gives immersive technologies the opportunities to change one's self [151].

For these reasons, we believe it is time to coin, at the terminological level, a new perspective that relates the two listening experiences (i.e. real and virtual), called **egocentric audio perspective**. In particular, we refer to the term audio to identify an auditory sensory component, implicitly recalling those technologies capable of immersive and interactive rendering. The term egocentric refers to the perceptual reference system for the acquisition of multi-sensory information in immersive VR technologies as well as the sense of subjectivity and perceptual/cognitive individuality that shape the self, identity, or consciousness. In accordance with Husserl's phenomenology, the human body can be philosophically defined as a "Leib" , a living body, and a "Nullpunkt" , a zero-point of reference and orientation [73].

This perspective aims to extend the discipline of Sonic Interaction Design [44] by taking into account not only the importance of sound as the main channel conveying information, meaning, aesthetic, and emotional qualities, but rather an egocentric perspective of entanglement between the perceiving subject and the computer simulating the perceived environment. In the first instance, this can be described by processes of personalization, adaptation, and mutual relations to maintain the immersive illusion. However, we will try to argue in this chapter that it is much more than that. We hope that our vision will guide the development of new immersive audio technologies and conscious use of sound design within VEs.

The starting point of this theoretical framework is an ecologically egocentric perspective. The foundational phenomenological assumption considers a self-propelled entity with agency and intentionality [47]. It can interact with the VE being aware of its activities in a three-dimensional space. The active immersion in a simulated acoustic field provides to it meaningful experiences through sound.

Therefore, It is important to introduce a terminological characterization of what is the **listener**, not a user in this context, as a human being with prior experience and subjective auditory perception. A closely related entity is the **auditory digital twin**, which differs from the most common avatar. The idea of an avatar within a digital simulation co-located with objects, places, and other avatars [126] requires a user taking control of any form of virtual bodies which might be noticeably different from that of the listener's physical body. On the other hand, the digital twin cannot disregard an egocentric perspective of the listener for whom it is created. This means that the relations with the VEs should consider personalization techniques on the virtual body closely linked to the listener's biological body. This mediation is essential for the interactions between the listener and all the diegetic sounds, whether they are produced by the avatar's gestures or by sound sources in the VE.

In such a context, immersiveness is a dynamic relationship between physical and meaningful actions by the listener in the VE. Specifically, having performed bodily practices such as walking, sitting, talking, grasping, etc. provide meaning to virtual places, objects, and avatars [59]. Accordingly, the sense of embodiment can be considered a subjective internal feeling which is an expression of the relationship between one's self and such VE. In this regard, Kilteni *et al.* [80] identified the sense of embodiment for an artificial body (i.e., avatar) in the mediation between the avatar's properties and their processing by the user's biological properties.

We now introduce the technological mediation in the form of an auditory digital twin which is a guardian and facilitator of (i) the sense of self-location, (ii) plausibility, (iii) body ownership, and (iv) agency for the listener. In the first instance, a performative view might make us see realities as "a doing", enacting practical actions [6, 104]. Similarly, the listener and the avatar cannot be considered fixed and independent interacting entities, but constituent parts of emergent, multiple and dynamic phenomena resulting from entangled social, cognitive, and perceptual elements. This **intra-systemic action** of entangled elements dynamically constructs identities and properties of the immersive listening experience. The illusory permanence of auditory immersion lies in the boundaries between situationally-entangled elements in fluid and dynamic situations. They can be seen as confrontations occurring exactly in the auditory digital twin that facilitates the phenomenon. The auditory digital twin is the meeting and shared place between the listener and a virtual body identity, communicating in a non-discursive (performative) way according to the quality level of the digital simulation.

In an immersive VE, the listeners cannot exist without their auditory digital twin and vice versa. Through the digital twin characterization, the acoustic signals generated by the VE are filtered exclusively for the listeners, according to their ability to extract meaningful information. It is worthwhile to mention the **participatory nature** of such entanglement process between listener and digital twin, as a joint exploration of the listener's attentional process in selecting meaningful information, e.g., the cocktails party effect [20]. We might speculate by considering a simulation that interacts within the digital twin to provide the best pattern or to discover it in order to attract the listener's attention. The decision-making process will then be the result of intra-action in and of the auditory digital twin.

This chapter has three main sections. Sec. 1.2 gathers the different souls that characterize the research and artistic works in SIVE. Section 1.3 holds a central position by defining the constitutive elements of our proposed egocentric audio perspective in SIVE: spatial centrality and entanglement between human and computer in the digital twin. In Sec. 1.4, we attempt incorporating this theoretical framework by adapting Milgram and Kishino's well-known taxonomy for VR [95], with an audio-first perspective. Finally, Sec. 1.5 concludes this chapter by encouraging a new starting point for SIVE. We suggest an inclusive approach to the next paradigm shift in the field of human-computer interaction (HCI) discipline.

## 1.2 SIVE: from an Archipelago to a Research Field

This chapter aims to provide an interpretation to an archipelago of researches from different communities such as:

- Sound and Music Computing (SMC) network, a point of convergence for different research disciplines mainly related to digital processing of musical information.[2]
- International Community for Auditory Display (ICAD), a point of convergence for different areas of research with digital processing of non-musical audio information and the idea of sonification in common.[3]
- The Audio Engineering Society (AES), the main community for institutions and companies devoted to the world of audio technologies.[4]
- The research community gathered by the International Conference on New Interfaces for Musical Expression (NIME), devoted to interactions with new interfaces with the aim at facilitating the human creative process.[5]
- The Digital Audio Effects community (DAFX) aiming at designing technological based simulations of sonic phenomena.[6]

We employ here the metaphor of an archipelago because it well describes a context in which all these communities address aspects of VR according to their specificities, influencing each others. After all, they share the same "waters". They are relatively close to each other but feeling distant from a VR community at the same time, like the islands of an archipelago in the open sea. Thus, we affirm the need to unify the fragmentary and specificity of those studies and to fill the gap with their visual counterpart's aiming at developing immersive VR environments for sound and music. To achieve this goal, the editors have pursued the following spontaneous path that is characterized by three main steps.

1. The first review article related to SIVE topics, dated back to 2018 [128], focused on the technological components characterizing an immersive potential for interactive sound environments. In that work, the editors and their collaborators produced a first compact survey including sound synthesis, propagation, rendering, and reproduction with a focus on the ongoing development of headphone technologies.
2. Two years later, we published a second review paper together with all the organizers of the past five editions of the IEEE Virtual Reality's SIVE workshop [129]. In this paper, we analyzed the contributions presented at the various editions highlighting the emerging aspects of interaction design, presence, and evaluation. An inductive approach was adopted, supported by a posteriori analysis of the characterizing categories of SIVE so far.

[2] https://smcnetwork.org/
[3] https://icad.org/
[4] https://www.aes.org/
[5] https://nime.org
[6] https://www.dafx.de/

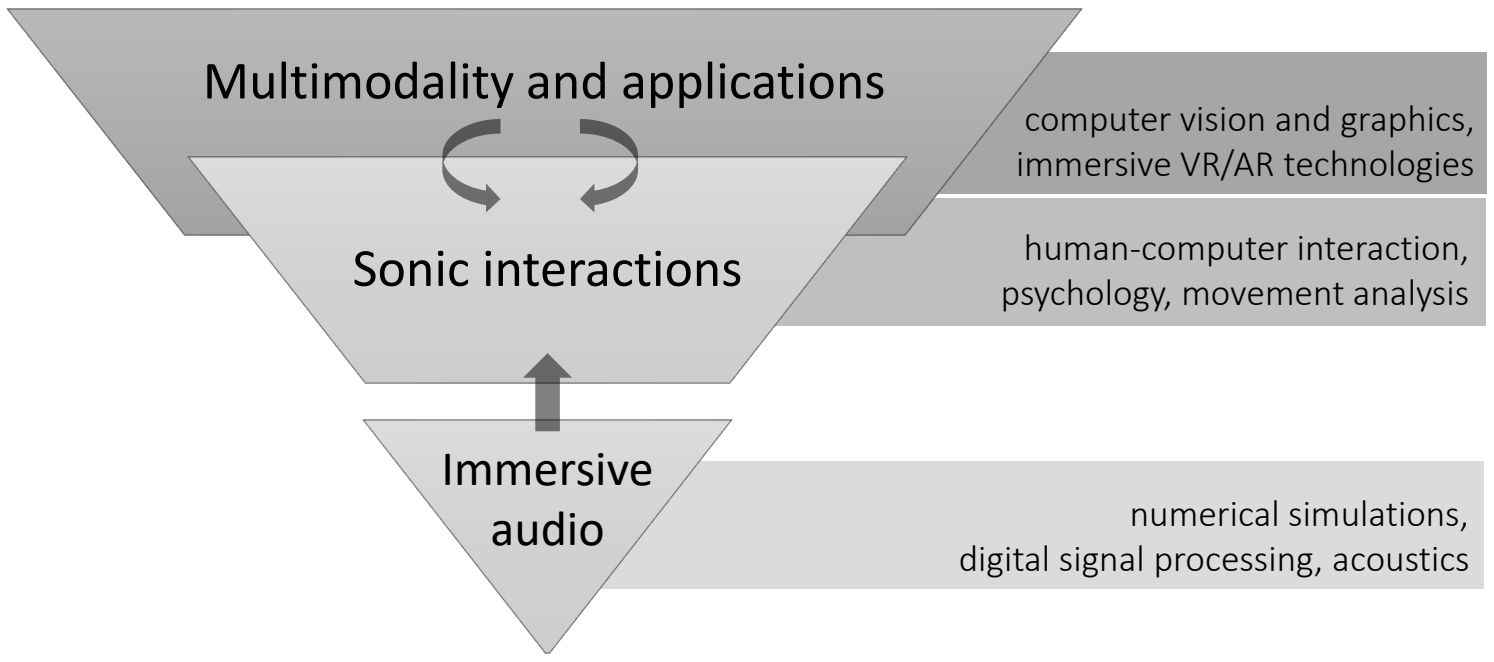

Fig. 1.1: The SIVE inverse pyramid. Arrows indicate high-level relational hierarchies.

3. Finally, this book and, in particular, this chapter want to raise the bar further with an organic and structured narrative of an emerging discipline. We aim to provide a theoretical framework for interpreting and accompanying the evolution of SIVE, focusing on the close relationship between physically-real and virtual auditory experiences described in terms of immersive, coherent, and entangled features.

This chapter is the result of the convergence of two complementary analytical strategies: (i) a *top-down* approach describing the structure given by the editors to the book originated from the studies experienced by the editors themselves, and (ii) a *bottom-up* approach drawing on the knowledgeable insights of the contributing authors of this book on several specialist and interdisciplinary aspects. Consequently, we will constantly refer to these chapters in an attempt to provide a unified and long-term vision for SIVE.

Our proposal for the definition of a new research field starts from a simple layer structure without claiming to be exhaustive. The graphical representation in Fig. 1.1 is capable of giving an overview and a rough inter-relation of the multidisciplinarity involved in SIVE. We suggest a hierarchical structure for the various disciplines in the form of an inverted pyramid representation. SIVE research can be conceptually organized in three levels:

i **Immersive audio** concerns the computational aspects of the acoustical-space properties of technologies. It involves the study of acoustic aspects, psychoacoustic, computational, and algorithmic representation of the auditory information and the development of enabling audio technologies;
ii **Sonic interaction** refers to human-computer interplay through auditory feedback in 3D environments. It comprises to the study of vibroacoustic information and its interaction with the user to provide abstract meanings, specific indicators of the state for a process or activity in interactive contexts;

iii The **integration** of immersive audio in multimodal VR/AR systems impacts different application domains. This third and final level collects all the studies regarding the integration of virtual environments in different application domains such as rehabilitation, health, psychology, music, to name but a few.

The immersive audio layer is a strongly characterizing element of SIVE. For such a reason, it is placed as the tip of the inverse pyramid, where all SIVE development opportunities originate. In other words, SIVE cannot exist without sound spatialization technologies, and the research built upon them is intrinsically conditioned by the level of technological development (for more arguments of this issue see Sec. 1.3.2).

In particular, spatial audio rendering through headphones involves the computation of binaural room impulse responses (BRIRs) to capture/render sound sources in space (see Fig. 1.2). BRIRs can be separated into two separate components: the room impulse response (RIR), which defines room acoustic properties, and the head-related impulse response (HRIR) or head-related transfer function (HRTF, i.e. the HRIR in the frequency domain), which acoustically describes the individual contributions of the listener's head, pinna, torso, and shoulders. The former describes the acoustic space and environment, while the latter converts this information into perceptually relevant spatial acoustic cues for the auditory system, taking advantage of the flexibility of **immersive binaural synthesis through headphones** and state-of-the-art consumer head-mounted displays (HMDs) for VR. The perceptually coherent auralization with lifelike acoustic phenomena, taking into account the ef-fects of near-field acoustics and listener specificity in user and headphones acoustics, is a key technological matter here [11, 21, 68].

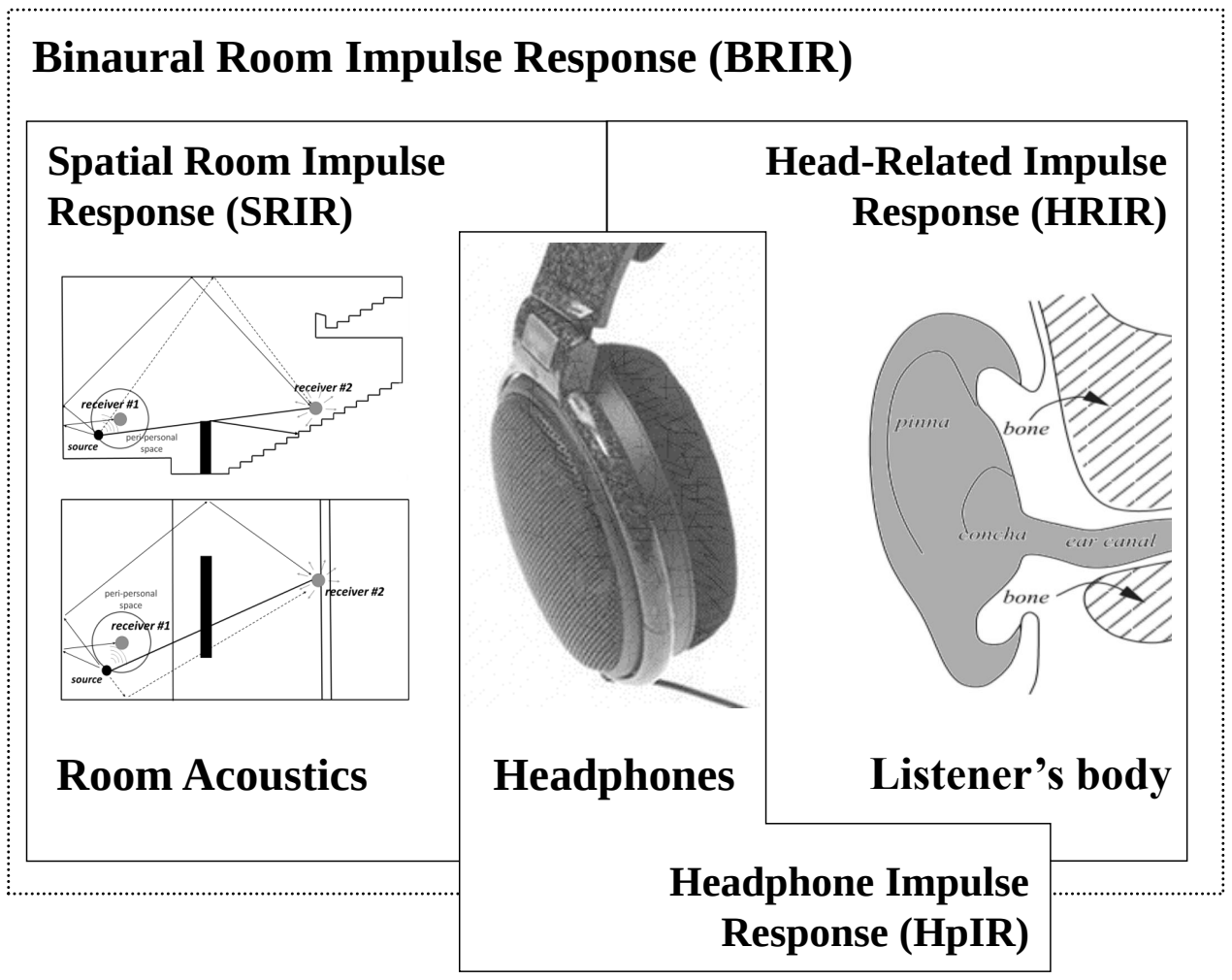

Fig. 1.2: High-level acoustic components for immersive audio with a focus on spatial room acoustics and headphone reproduction.

The visual component of spatial immersion is so evident that it may seem that the sensation of immersion is exclusively dependent on it, but the aural aspect has as much or even more relevance. We can simulate an interactive listening experience within VR using standard components such as headsets, digital signal processors (DSPs), inertial sensors, and handheld controllers. Immersive audio technologies have the potential to revolutionize the way we interact socially within VR environments and applications. Users can navigate immersive content employing head motions and translations in 3D space with 6 degrees of freedom (DoF). When immersive auditory feedback is provided in an ecologically valid interactive multisensory experience, a perceptually plausible scheme for developing sonic interactions is practically convenient [128], yet still efficient in computational power, memory, and latency (refer to Ch. 3 for further details). The trade-off between accuracy and plausibility is complex and finding algorithms that can parameterize sound rendering remains challenging [62]. The creation of an immersive sonic experience requires:

- *Action sounds*: sound produced by the listener that changes with movement
- *Environmental sounds*: sounds produced by objects in the environment, referred to as soundscapes
- *Sound propagation*: acoustic simulation of the space, i.e. room acoustics
- *Binaural rendering*: user-specific acoustics that provides for auditory localization

These are the virtual acoustics and auralization key elements [153] at the basis of auditory feedback design that draws on user attention and enhances the sensation of place and space in virtual reality scenarios [102].

The two upper layers of the SIVE inverse pyramid, i.e. sonic interactions and multimodal experiences, are not clearly distinguishable and we propose the following interpretation: we differentiate the interaction from the experience layer when we intend to extrapolate design rules for the sonic component with a different meaning for the designer, system, users, etc. . In both cases, embodiment and proprioception are essential, naturally supporting multimodality in the VR presence. This leads us to a certain difficulty in generalizations which is well-grounded by our egocentric audio perspective. In our proposal of theoretical framework, the hierarchies initially identified can change dynamically.

Ernst and Bülthoff's theory [41] suggests how our brain combines and merges different sources of sensory information. The authors described two main strategies: sensory combination and integration. The former aims at maximizing the information extraction from each modality in a non-redundant manner. The second aims at finding congruence and reducing variability in the redundant sensory information in search of greater perceptual reliability. Both strategies consider a bottom-up approach to sensory integration. In particular, the concept of *dominance* is associated with perceptual reliability from each specific sensory modality given the specific stimulus. This means that the main research challenge for SIVE is not only to foster research aimed at understanding how humans process information from different sensory channels (psychophysics and neuroscience domains), but especially how multimodal VEs should distribute the information load to obtain the best experience for each individual. Accordingly, we assume that each listener has **personal optimization strategies** to extract meaning from redundant sensory information distributions. The VR technology can improve if and only if it can have a sort of dialogue with the listener to understand such a natural mixture of information .

The design process of multimodal VEs must also constantly take into account the limitations, i.e. the characterization, of the VR technologies with the aim at creating real-time interactions with the listener. According to Pai [108], interaction models can be described as a trade-off between accuracy and responsiveness. Increasing the descriptive power and thus the accuracy of a model for a certain phenomenon leads to processing more information before providing an output in response to a parametric configuration. It comes at the price of higher latency for the system. For **multisensory models** that should synchronize different sensory channels, this is crucial and has to be carefully balanced with many other concurrent goals.

Understanding interactions between humans and their everyday physical world should not only inspire the design of natural multimodal interfaces but should be directly explored into VE models and simulation algorithms. This message is strongly supported by Ch. 10 and our theoretical framework fully integrates this vision by trying to further extend this perspective to non-human agents. The role of the digital simulation and the computer behind it is participation and discovery for the listener. They constitute a complex system whose interactions contribute to the dynamic definition of non-linear narratives and causal relationships that are crucial for immersive experiences. The application contexts of the interactive simulations instruct the tradeoff between the accuracy and responsiveness models. Hence, the knowledge of the perceptual-cognitive listener capabilities emerges as active transformations in multimodal digital VR experiences.

## 1.3 Egocentric Audio

A large body of research in computational acoustics focused on the technical challenges of quantitative accuracy characterizing engineering applications, simulations for acoustic design, and treatment in concert halls. Such simulations are very expensive in terms of computational resources and memory, so it is not surprising that the central role of perception in rendering has gradually come into play. The search for lower bounds such as the perceptually authentic audio-visual renderings can be achieved (see Ch. 5 for a more detailed discussion). Continuous knowledge exchange between psychophysical research and interactive algorithms development allows to test new hypotheses and propose responsive VR solutions. It is worthwhile to mention the topic of artificial reverberations and modeling of the reverberation time aiming to provide a sense of presence through the main spatial qualities of a room, e.g. its size [83, 147].

In the context of SIVE, we could review and adapt the three paradigm shifts, or "waves" in HCI mentioned by Harrison [64], which still coexist and are at the center of research agendas for different scientific communities. The first wave considers the optimization of interaction in terms of the human factor in an engineered system. We could mention as an example the ergonomic, but generic, " one fits all" solutions of dummy-heads and binaural microphones for capturing acoustic scenes [110]. The second wave introduces a connection between man and machine in terms of information exchange, looking for similarities and common ground in decision-making processes, e.g. memory and cognition. The structural inclusion of non-linearities and auditory Just-Noticeable Differences (JNDs) to determine the amount of information to be encoded for gesture sonification is an example of this direction [38]. Finally, the third paradigm shift considers interaction as a situated, embodied, and social experience, characterized by emotions and complex relations encountered in everyday life. We could place here many of the case studies collected in this volume (Parts III and IV). To this regard, the extracted patterns or best practices are often very specific to each study and listeners' groups, e.g. musician vs. non-musician (Ch. 9).

From developments in phenomenological [93] and, more recently, post-phenomenological thinking by [74, 150], we will therefore develop the egocentric audio perspective. The key principle is the shift between inter-action between defined objects to intra-action within a phenomenon whose main actors are human and non-human agents. Boundaries between actors are fluidly determined, similarly to the Gibsonian ecological theory of perception [54, 55]. Even though this is a shift from an anthropocentric and user-centered view towards a system of enactive relations and associations in the immersive world of sounds, we chose the term egocentric to emphasize the **spatial anchoring** between humans and technology in the self-knowledge constitution.

It would be useful also referring to the concept of *ambiguity* by the philosopher Maurice Merleau-Ponty that says that all experiences are ambiguous, composed of things which do not have defined, identifiable essence, but rather by open or flexible styles or patterns of interactions and developments [93, 123]. Starting from an egocentric spatial perspective of immersive VR, the learning and transformation processes of the listeners occur when their attention is guided towards external virtual sounds, e.g. the out-of-the-head and externalized stimuli. This allows them to achieve meaningful discoveries also for their auditory digital twins. Accordingly, the experience mediated by a non-self, i.e. auditory simulation of VEs, is shaped (i) by the past experience of the listener and the digital twin indistinctly acquired from a physical or cybernetic world in a constructivist sense, (ii) by the physical-acoustic imprinting induced or simulated by the body, head and ears, and (iii) by active and adaptive processes of perceptual re-learning [57, 160] induced by **a symbiosis with technology**. Figure 1.3 schematizes and simplifies this relationship between man-technology-world from which the listener acquires meaning. As pointed out by Vindenes and Wasson [151], experiences are mediated in a situated way from the subjectivity of the listener which constitutes herself in relation to the objectivity of the VE. Having placed the physical and virtual worlds at the same level yields to similar internal representations for the listener and her digital twin, allowing us to promote the transformative role of VR experiences for a human-reality relationship altered after exposure.

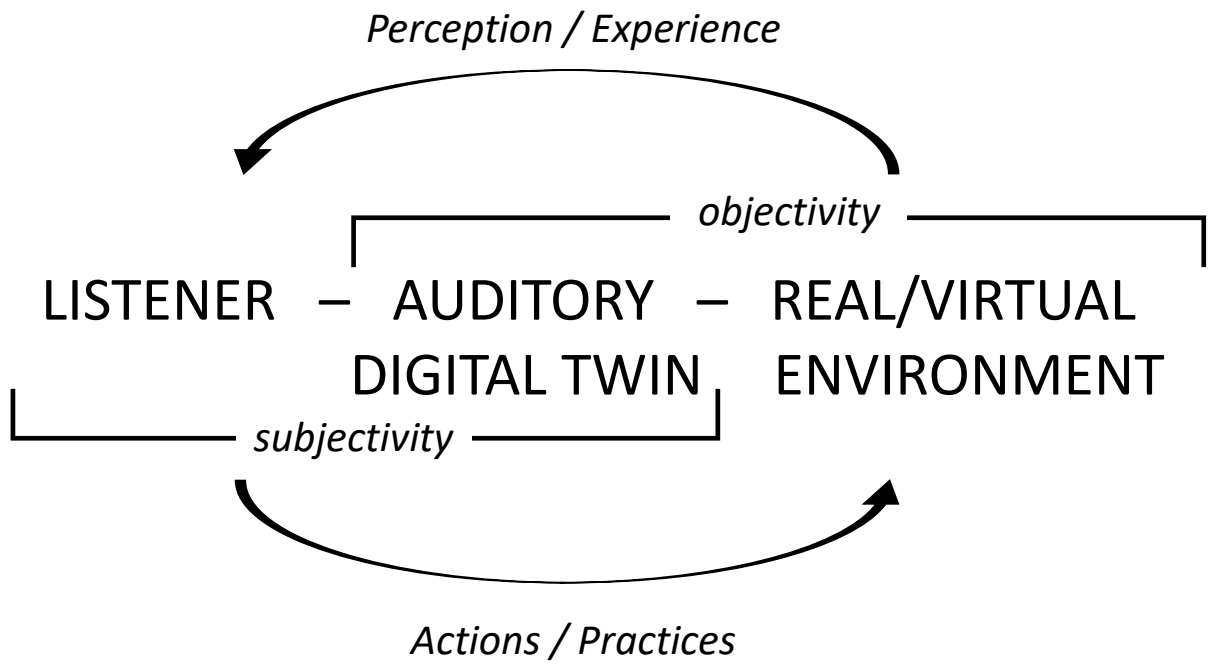

Fig. 1.3: Technological mediation of the auditory digital twin (adapted from Hauser *et al.* [66]).

The core of our framework is an ideal **auditory digital twin**: an essential mediator and existential mirror for an egocentric audio perspective. Technology is the mediator of this intentional relationship co-constituting both the listener and her being in the world. From this post-phenomenological perspective of SIVE, we are interested in understanding how the VE relates to the listener and what is the meaning of the VEs for the listener, at the same time. Our main goal is to characterize the mediating action between the listener and the VE by an auditory digital twin. This guardian can reveal the listener's ongoing reconfiguration through the human-world relationship occurring outside the VR experience.

In the remainder of this chapter, we will motivate the opportunity to refer to this non-human entity other than the self and aspiring to be the mediator for the self. This first philosophical excursus of hermeneutical nature allows us to take a forward-looking vision for the SIVE discipline, framing the current state of the art but also including the rapid technological developments and ethical challenges due to the digital transformation.

### 1.3.1 Spatial Centrality

The three-dimensionality of the action space is one of the founding characteristics of immersive VE. Considering such space of transmission, propagation, and reception of virtually simulated sounds, sonic experiences can assume different meanings and open up to many opportunities.

Immersive audio in VR can be reproduced both through headphones and loudspeaker arrays determining a differentiation between listener- and loudspeaker-centric perspectives. The latter seems to decentralize the listener role in favor of a strong correlation between virtual and physical (playback) space. In particular, sound in VEs is decoded for the specific loudspeaker arrangements in the physical world (for a summary of the playback systems refer to Ch. 5). This setup allows the coexistence of several listeners in the controlled playback space, depending on the so-called sweet spot. However, the VE and the listener-avatar mapping is intrinsically egocentric and multisensory, subordinating a loudspeaker-centric perspective for the simulation of the auditory field to a listener-centric one. Let us try to clarify this idea with a practical example: head movements and the navigation system e.g. redirected walking [101], determine the spatial reference changes for the real/virtual environment mapping corresponding to the listener's dynamic exploration. The tracking system could trigger certain decisions to maintain the place and plausibility illusions of the immersive audio experience.

#### 1.3.1.1 First Person Point of View

In this theoretical framework, we focus on the listener's perspective, where sound is generated from the first-person point of view (generally referred to as 1PP). Virtual sounds are shaped by spatial hearing models: auralization takes into account the individual everyday listening experience both in physical-acoustic and non-acoustic terms. Contextual information relate spatial positions between sound events and objects with the avatar virtual body, creating a sense of proximity and meaningful relations for the listener.

It is relevant to stress the connection between the egocentric audio perspective and the research field of egocentric vision that has more than twenty-year history. The latter is a subfield of computer vision that involves the analysis of images and videos captured by wearable cameras, e.g. Narrative Clips[7] and GoPro[8], considering an approximation of the visual field due to a 1PP. From this source of information, spatio-temporal visual features can be extracted to conduct various types of recognition tasks, e.g. of objects or activities [100], and analysis of social interactions [2]. The egocentric audio perspective originates from the same 1PP in which both space and time of events play a fundamental role in the analysis and synthesis of sonic interactions. Furthermore, we stress the idea that all hypotheses and evaluations in both egocentric vision and audition are individually shaped around a human actor. However, our vision does not focus exclusively on the analysis of the listener behaviours but includes **generative aspects** thanks to the technological mediation of the spatial relations between humans and VEs (these aspects will be extensively discussed in Sec. 1.3.2).

Using a simplification adopted in Ch. 2 concerning the work by Stockburger [140] on sounds in video games, we can distinguish two categories for sound effects: (i) those related to the avatar's movements and actions (e.g. footsteps, knocking on a door, clothing noises, etc.) and (ii) the remaining effects produced by the VE. In this simple distinction, it is important to note that all events are echoic, i.e., they produce delays and resonances imprinted by the spatial arrangements of the avatar-VE configurations depending on the acoustical characteristics of the simulated space. Moreover, all events should be interpreted by the listener's perceptual memory which is shaped by the natural everyday reality.

[7] http://getnarrative.com/

[8] https://gopro.com/

Finally, it is worthwhile to notice that egocentric 1PP poses novel challenges in the field of cinematic VR narration or more generally of storytelling in VR. Gödde *et al.* [56] identified immersive audio as an essential element able to capture attention on events/objects outside the field of view. The distinction between the active role of the listener interacting with the narrative or passive role as an observer raises interesting questions about the spatial and temporal positioning of scenic elements. The balance between environment, action, and narration is delicate. Citing Gödde and collaborators, one *"can only follow a narrative sufficiently when temporal and spatial story density are aligned with each other."*. Hence, the spatio-temporal alignment of sound is crucial.

For most researchers interested in sound, from the neurological to the aesthetic-communicative level, it is clear that while the visual object exists primarily in space, the auditory stimulus occurs in time. Therefore, it is not surprising that in order to speak of spatial centrality in audio we need to consider presence, the central attribute for a VR experience. In his support of a representational view of it, Loomis [88] cites two scientists with two opposite opinions: Willian Warren and Pavel Zahorik, the first an expert in visual VR and the latter in acoustic VR. The former supports a rationalist view of representational realism and direct perception [154], while the latter supports the ecological perspective in the **fluidity in perception-action** [159].[9] The second perspective supports the concept of enaction such that it is impossible to separate perception from action in a systematic way. Perception is inherently active and reflexive in the self. Recalling Varela, another leading supporter of this perspective [148], experience does not happen within the listener but is instead enacted by the listener by exploring the environment. Accordingly, we consider an embodied, environmentally situated perceiver where sensory and motor processes are inseparable from the exploratory action in space. At first glance, such a view restricts experiences to only those generated by specific motor skills which are in turn induced by biological, psychological, and cultural context. However, it is generally not true in a digital-twin-driven VE (see Sec. 1.4.3).

#### 1.3.1.2 Binaural Hearing

The geometric and material features of the environment are constituent elements of the virtual world that must be simulated in a plausible way for that specific listener. First of all, the listener-environment coupling is unavoidable and must guarantee as good sound localization performances as to maintain immersiveness. It has to espe-cially avoid the inside-the-head spatial collapse, i.e. when the virtual sound stimuli are perceived inside the head, a condition opposite to the natural listening experience of outside-the-head localization for surrounding sound sources, also called externalization [131]. Externalization can be considered a necessary but not sufficient condition for the place illusion, being immersed in that virtual acoustic space. For a recent review of the literature on this topic, Best *et al.* [8] suggest that ambient reverberation and sensorimotor contingencies are key indicators for eliciting a sense of externalization, whereas HRTF personalization and consistent visual information may reinforce the illusion under specific circumstances. However, the intra-action between these factors is so complex that **no univocal priority principles** can be applied. Accordingly, we should explore dynamic relations depending on specific links between evolving states of the listener-VE system during the VR experiences. Moreover, huge individual-based differences in the perception of externalization require in-depth exploration of several individual factors such as monaural and binaural HRTF spectral features, temporal processes of adaptation [27, 65, 146].

[9] Atherton and Wang [4] recently developed a similar view point comparison and proposed a set of design principles for VR, born from the contrast between "doing vs, being" .

Binaural audio and spatial hearing have been well-established research fields for more than 100 years and have received relevant contributions from information and communications technologies (ICT) and in particular from digital signal processing. Progress in digital simulations has made it possible to replicate with increasing accuracy the acoustic transformation by the body of a specific listener with very high spatial resolution up to sub-millimeter grids for the outer ear [113, 114]. This process generates acoustically personalized HRTFs so that the rendering of immersive audio matches the listener's acoustic characterization (*System-to-User adaptation* in Ch. 4). On the opposite side, the VE can train and guide the listener in a process of *User-to-System adaptation* by designing ad-hoc procedures for continuous interaction with the VE to induce a persistent **recalibration of the auditory system** to non-individual HRTFs.[10] These two approaches can be considered two poles between which one can define several mixed solutions. This dualism is brilliantly exposed and analyzed in Ch. 4.

#### 1.3.1.3 Quality of the Mediated Experience

Since our theoretical framework aims to go beyond user-centricity, we approach the space issue from different perspectives, both user and technology perspectives, respectively. However, all points of view remain ecologically anchored to the egocentric 1PP of the listener giving rise to a fundamental question: how can we obtain high-quality sonic interactions for a specific listener-technology relation? In principle, many **quality assessment procedures** might be applied to immersive VR systems. However, there is no adequately in-depth knowledge of the technical-psychological-cognitive relationship regarding spatial hearing and multisensory integration processes linked to plausibility and technological mediation.

---

[10] The HRTF selection process can potentially result from a random choice [139].

On the other hand, a good level of standardization has been achieved for the perceptual evaluation of audio systems. For instance, the ITU recommendations focus on the technical properties of the system and signal processing algorithms. Chapter 5 introduces the *Basic Audio Qualities* used for telecommunications and audio codecs, commonly adopted in the evaluation of spatial audio reproduction systems. On the other hand, the evaluation of the listening experience quality, called *Overall Listening Experience* [125], is also introduced, considering not only system technical performances but also listeners' expectations, personality, and their current state. All these factors influence the listening of specific audio content. A related measure can be the level of audio detail (LOAD) [39] that attempts to manage the available computational power, the variation of spatio-temporal auditory resolution in complex scenes, and the perceptual outcome expected by the listener, in a dynamically adaptive way.

Chapter 2 provides an original discussion on audio "quality scaling" in VR simulations, drawing the following conclusion: there is neither an unambiguous definition nor established models for such issues. It suggests that understanding the listener-simulation-playback relations is an open challenge, extremely relevant to SIVE. In general, the most commonly used approach is the **differential diagnosis**, allowing the qualities of VR systems to emerge. Different taxonomies for audio qualities or sound spatialization have given rise to several attribute collections, e.g. semantic analysis of expert surveys and expert focus groups (see Ch. 5 on this). It is worthwhile to mention that a substantial body of research in VR is devoted to explore the connections between VR properties such as authenticity, immersion, sense of presence and neurophysiological measurements, e.g. electroencephalogram, electromyography, electrocardiogram, and behavioral measurements, e.g. reaction time, kinematic analysis.

To summarize, this differentiation tries to capture all those factors that lead to a high level of presence: sensory plausibility, naturalness in the interactions, meaning and relevance of the scene, etc. . Moreover, the sense of presence in a VR will remain limited if the experience is irrelevant to the listener. If the listener-environment relation is weak, the mediating action of the immersive technology might result in a break in presence that can hardly be restored after a pause [136]. These cognitive illusions depend, for example, on the level of hearing training, familiarity with a stimulus/sound environment. All these aspects reinforce the term egocentric again, grounding auditory information to a reference system that is naturally processed and interpreted in 1PP. However, SIVE challenges go far beyond two opposing points of view, i.e. user-centered and technology-centered. In this chapter, we offer a first attempt at a systemic interpretation of the phenomenon.

### 1.3.2 Entanglement HCI

Heidegger's phenomenology aims to overcome mind-body dualism by introducing the notion of 'Dasein' which requires an embodied mind to be in the world [67].

The concept of embodiment thus became central to the third wave of HCI, e.g. in relation to mobile and tangible user interfaces [64]. More recently, the bodily element has been incorporated into the theoretical framework of somaesthetics to explain aesthetic experiences of interaction and into **design principles for bodily interaction** [71]. Designers are encouraged to participate with their lived, sentient, subjective, purposive bodies in the process of creating human-computer interactions, either by improving their design skills and sensibilities, or by providing an added value of aesthetic pleasure, lasting satisfaction, and enjoyment to users. These elements are summarized in Ch. 7, which provides a useful distinction of perspectives for interaction design: the 1st-person, 2nd-person, and 3rd person design perspective. The latter is equivalent to an observer approach to design such as considering the common practices, e.g. interview administration, subjective evaluations, and data analysis acquired from a variety of sensors. The 2nd-person is equivalent to the user-centered and co-design approach between the user's perspective and the designer's attempt to step into the shoes of someone else. On the other hand, soma design principles embrace a 1st-person perspective, we would argue egocentric, even for designers, who are actively involved with their bodies during each step of the interaction design process of an artifact or simulation. They explicitly become actors themselves with the result of shaping a felt and lived experience for other actors.

In the movement computing work by Loke and Robertson [87], the authors introduced another perspective distinction relevant here. The mover (1st-person perspective) and the observer (3rd-person perspective) are explicitly joined by **the machine perspective**. The role of technology is pivotal for the interactions with digital movement information and, in particular, for the process of attributing meaning based on user input. This perspective requires mapping data from sensing technologies into meaningful representations for the observer and the mover. It is worthwhile to note that machines capture the qualities of movement with considerable losses in terms of spatial, temporal, or range resolution, making the comprehension of such limitations on interaction design essential. We need to explore the various perspectives, not in a mutually exclusive way, but dynamically managing the analysis of the various points of view in every immersive experience.

According to Verbeek [150], human-world relations are enacted through technology. Thus, man and technology constitute themselves as actors in a **fluid reconfiguration of themselves**. A practical example in the field of music perception considers a drummer who changes her latency perception the more she plays the musical instrument [86]. The action of playing the drum changes the relationships that she has with the instrument itself, with the self, and with temporal aspects of the world, e.g. reaction times, synchronizations.

The recent proposal of a post-phenomenological framework by Vindenes [151] is based on Verbeek's concept of technological mediation, which identifies several human-technology relationships including immersion in smart environments, ambient intelligence, or persuasive technologies. In particular, for the latter case, VR plays a central role co-participating within a mixed intentionality between humans and technology. Accordingly, Verbeek introduced the idea of composite intentionality for cyborgs [149], a cooperation between human and technological intentionality with the aim to reveal a (virtual) reality that can only be experienced by technologies, by **making accessible technological intentionalities to human intentionality**. We can argue that the world and the technology become one in the immersive simulation that knows the listeners and actively interacts with them. This configuration becomes bidirectional: humans are directed towards technology and technology is directed towards them. Moreover, listeners have the opportunity to access reflective relationships to themselves through VEs. For example, Osimo *et al.* provided experience of the self through virtual body-swapping in the embodied perspective-taking [106]. We must decentralize humans as the sole source of activity and attribute to the material/technological world an active role in revealing new and unprecedented relational actions.

This approach opens up new opportunities for "reflexive intentionality" of the human beings about themselves through the active relation with simulations [5]. About this, Verbeek [150] classifies the technological influence on humans according to two dimensions: visibility and strength. Some mediations can be hidden but induce strong limitations, while others can be manifest but have a weak impact on humans. There is a deep entanglement between humans and machines to the extent that there is no human experience that is not mediated through some kind of technology that shapes who we are and what we do in the world. Considering immersive VR technologies, we must speculate on what is a **locus of agency**:the understanding of the active contributions of each tool in the listener's actions in VEs. Such an infrastructure must be enactive and re-interpretive of each actor in each circumstance. In other words, there is the opportunity of becoming different actors depending on an active inter-dependence.

At this point, recalling the work of Orlikowski [105] is twofold. First, she gave the name of *entanglement* theories to those heterogeneous theories that have in common the recognition of the active inter-dependence between socio-technological-material configurations with the consequence of promoting studies of man and technology in a unitary way. Secondly, Orlikowski supported her position with an experimental example of social VR, the Sun Microsystems' Project Wonderland developed more than a decade ago and, nowadays, it seems more relevant than ever due to the COVID-19 pandemic. We will analyze a similar case in SIVE, supporting our taxonomy in Sec. 1.4. In this section, we focus on entanglement theories that are foundational for our egocentric perspective.

The entanglement is the deep connection between men and their tools, having relevant repercussions in the field of human-computer interaction. In [45], Frauenberger provided the following interpretative key: we cannot design computers or interaction design, we can work on facilitating certain configurations that enact certain phenomena. Both configurations and phenomena are situated and fluid, but not random. They are causally connected within **hybrid networks in which human and non-human actors interact**. However, it must be made clear that these actors do not possess fixed representations of their entities, but they exist only in their situated intra-action. This means that their relations and configurations are dynamically defined by the so-called *agential cuts* that draw the boundaries between entities during phenomena. In this network of associations, each configuration change is equivalent to a newly enacted phenomenon where new agential cuts are redefined or create new actors. Hence, the term agency refers to a performative mechanism of boundary definition and constitution of the self. Together with the post-phenomenological notion of technological mediation, entangled HCI provides a lens able to interpret the increasingly fuzzy boundaries between humans, machines, and their distribution of agency.

The sonic information from intentional active listening is anchored to an egocentric perspective of spatiality that allows the understanding of an acoustic scene transformed by the listener's actions/movements. This process can be mathematically formalized with the active inference approach by Karl Friston and colleagues [46] and their recent enactive interpretation [115]. Their computational framework quantitatively integrates sensation and prediction through probability and generative models optimizing the so-called *free-energy principle*, i.e., an optimization problem of a function of the beliefs and expectations. Following this line of thought both philosophically and mathematically, we argue that immersive audio technologies are capable of contributing to the listener's internal representation in both spatial and semantic terms, eliciting a strong sense of presence in VR [12]. Just as we cannot clearly distinguish between listener and real environment, the more we cannot distinguish between listener and VE.

Therefore, the sonic interaction design in VEs is an intra-action between technology, concepts, visions, designers, and listeners that produce certain configurations and agential cuts. According to the sociological actor-network theory [28, 85], the network of associations characterizes the ways in which materials join together to generate themselves. Prior knowledge also becomes an actor in such a network that shapes, constrains, enables, or promotes certain activities. For example, modeling the listener's acoustic contribution with measurements from a dummy head induces a cut that shapes the use cases and VR experiences. Similarly, agential cuts are performed based on knowledge from other studies. For instance, the auditory feedback supports the plausibility of footstep synthesis or the strategies employed in the definition of time windows for synchronous and embodied sensory integration [122]. Moreover, the physical and design features of the technology also contribute to determining what is feasible: e.g., the differentiation of playback systems for spatial audio results in differentiation in the quality of the experience (see Ch. 11).

In the entanglement within the relational network of listener-reality-simulation, configurations and actors are dynamically defined in a situated and embodied manner. In the process of configuring and reconfiguring actors, designing various aspects, and operating agential cuts **new knowledge is produced** that causally links the enactment of the technological design to the phenomenon created [45]. This means that this knowledge has several forms, one resides in the technological artifact itself, i.e. in the VR simulation. In a more general sense, we could argue that exploring the evolution in the network configurations and actors enables an active search of the egocentrically meaningful experience. In line with this, agency and its responsibilities are not the prerogative of the listener or the technology but reside in their intra-actions.

### 1.3.3 Auditory Digital Twin

From entanglement theories, we inherit a series of open questions that guides our reflection on the SIVE research field. Let's consider the immersive VR simulation as the digital artifact co-defining itself with the listener who experiences it.

*How can certain transformative actions and interactions be programmed?*

*Who/what is the mediator, if any, in the relationship between the physical world and the VE?*

*How should it act?*

Of particular interest here is Schultze's interpretation of the avatar [126]: a dynamic self-representation for the user, a form of situated presence that is variably implemented. Sometimes the avatar is seen as a separate entity, behaving independently of the user. Sometimes the listener inhabits the avatar, merging with it to such an extent that they feel completely immersed and present in the virtual space. From this variety of instances, definitions of identity (avatar vs. self), agency (technology vs. human), and the world (physical vs. virtual) are fluid and enacted depending on the situation. Moreover, we argue that avatars and listeners know very little about each other. Such consideration strengthens the individual experience that determines one tendency over the other (separation vs. union with an avatar) with difficult predictions and poorly generalizable interpretations. Consequently, the user characterization in human-centered design is somehow included here [76]. However, our view promotes meaningful human-technology relationships in a bidirectional manner: not only personalized user experiences, but **experiences able to shape who we really want to be**.

The communication between the avatar and the listener, the virtual and the physical is challenging. Considering the avatar as part of a VE configuration, we can formulate one of the initial questions: if we can handle mediation, where/who is in charge of that?

Our performative perspective is questioning the a priori and fixed distinctions of certain representationalism between avatar and self, technology agency and listener, physical reality and virtuality. These boundaries have to be drawn in situated and embodied action, which makes them dynamic and temporary. The exploration of how, when, and why agential cuts define boundaries of identity, agency, and environments is the core of our theoretical framework.

We want to give a digital form to the philosophical question of the *locus of agency*: we envision a meta-environment with technological-digital nature, which is the guardian, careful observer, and lifeblood for the dialogue and participation of each actor. Its name is the **auditory digital twin**. In an egocentric perspective, it takes shape around the listener, i.e. the natural world that is meaningful to her. Why twin? Because this term recalls the idea of the deep connection between two different and distant entities or persons, commonly grounded by similarities, e.g. the DNA or a close friendship. Although the adjective auditory would seem to restrict our idea to the sound component, the framework ecologically extends to the multisensory domain by considering the intrinsic multisensory nature of VR. For these reasons, we will provide an audio-first perspective, sometimes sacrificing the term auditory in favor of a more readable and synthetic expression without loss of information, i.e. (auditory) digital twin.

Technical aspects of an artifact can be used to recreate a virtualized version or digital simulation of the artifact itself in the so-called virtual prototyping process [90]. Similarly, perceptual and cognitive aspects might serve to obtain digital replicas of biological systems, also referred to as a bio-digital twin in the field of personalized medicine [23]. The real person/machine provides the data that gives shape to the virtual one. In the case of humans, the process of **quantified self** [89] supports the modeling of the virtual digital twin, a personal assistant in decision-making. Implications of the digital twin paradigm are already envisioned in [40]. They range from the continuous monitoring of patient health to the management of the agency in a potentially immortal virtual agent.

In the scientific literature, the most common definition of a digital twin is related to a digital replica. However, we would like to provide a significant imprint to our idea of the auditory digital twin as a **psycho-socio-cultural-material objectified actor-network with agential participation**. As depicted in Fig. 1.4, all digitally objectifiable configurations related to listener profile, VE, HW/SW technology, design, ethical impact, etc. are made available to the digital twin so that it can actively participate intra-acting with system states.

To understand the central role of the digital twin in SIVE, we provide some practical examples:

- **Links to setup configurations** - Body movement tracking opens up numerous opportunities for dynamic rendering and customization of the listener's acoustic contribution in harmony between the real and the virtual body, i.e. the avatar's body. Real-time monitoring of the motion sensors is crucial to avoid a negative impact on responsiveness.
- **Links to listener configurations** - Adaptation and accommodation processes are strongly situated in the task. Assuming the unavailability of individual HRTF measurements, the best HRTF model requires a dynamic analysis of each task/context in a mutual learning perspective between the listener and the digital twin.
- **Links to environment configurations** - Persuasion of a VE for a listener behavioral change depends on social and cultural resonances within the listener. The distribution of agency in a music-induced mood has to be analyzed with particular attention. Again, certain immersive gaming experiences or role-playing may be beneficial for some listeners, to be avoided for others.
- **Links to configurations of others** - Other entities, e.g. virtual agents or avatars guided by other listeners, populate VEs. To manage confrontation and sharing activities, the intra-action between a larger number of digital twins must be consciously encouraged.

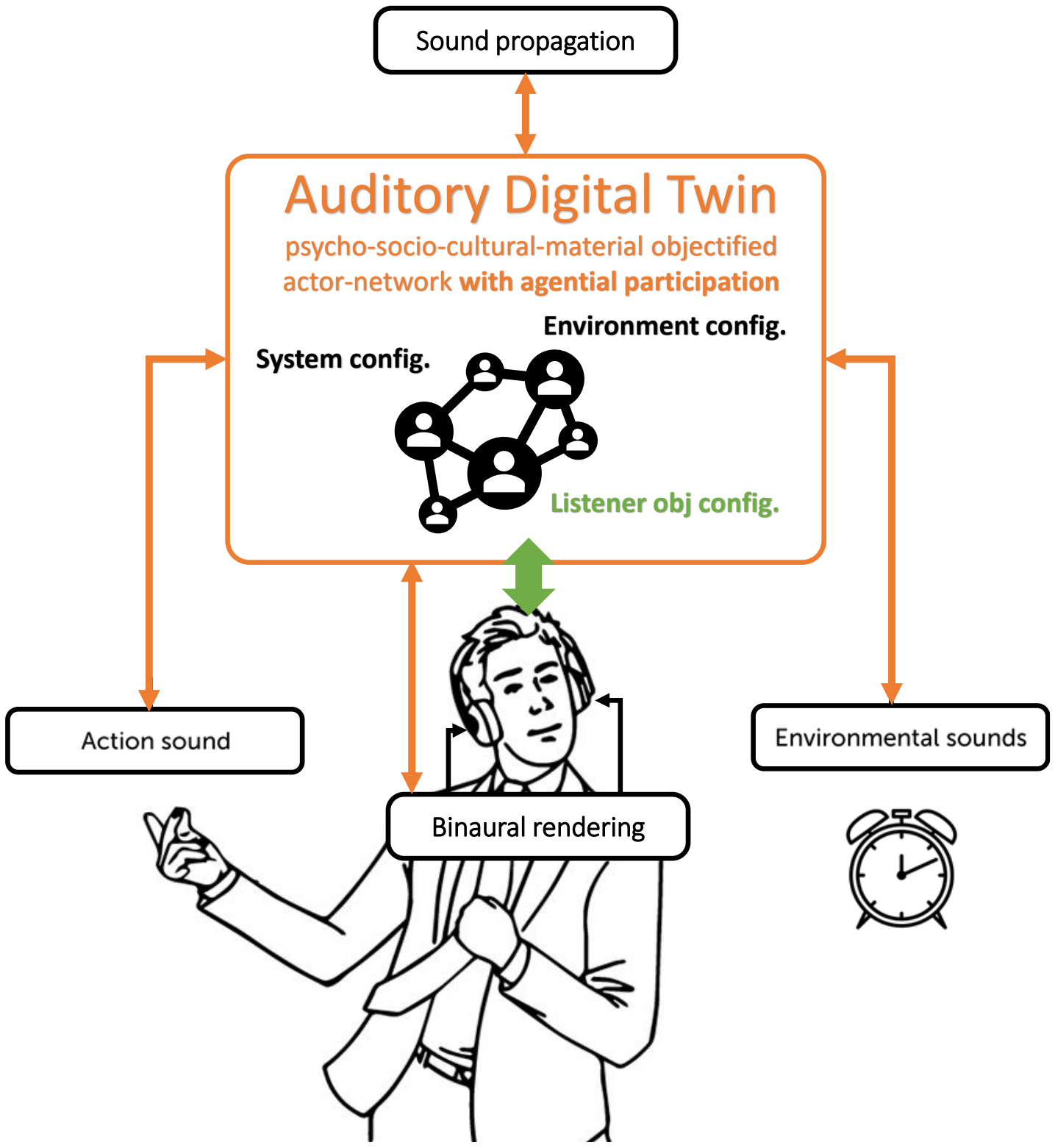

Fig. 1.4: A schematic representation of the different sound elements needed to create an immersive sonic experience. Colored lined identify the differences compared to the scheme proposed in [128]. In particular, this representation focuses on the central role of the auditory digital twin as a quantifiable *locus of agency* in an active relationship with all actors of a VR experience. The green arrow identifies the participatory relationship between the listener and the digital twin in its performative formation of individual self-knowledge.

All these configurations are not independent but are always interconnected with each other. Of particular relevance here, we can consider the externalization of sound sources. The level of externalization depends on customization techniques of the spatial audio rendering, the acoustic information of the virtual room, the sensory coherence and synchronicity, and the familiarity with the situation [8]. A coordination action of setup, environment, and listener(s) is needed. The presence in VR experiences will be the result of all these fluid intra-connections.

Suchman posed a highly relevant question in [141]: how can we consider all these configurations in such a way that we can act responsibly and productively with and through them? To answer, we must deal with the participation issues for all involved actors.

The egocentric perspective requires us to start from the listener and her experience. The scientific literature already tells us that memory, comprehension, and human performance benefit considerably from these VEs, especially in guided or supervised tasks involving human or digital agents [29]. Let us focus on the series of actions triggered by an active role of agents. In [31], Collins analyzed the player role in the audio design of video games. The participatory nature of video games potentially leads to the creation of additional or completely new meanings compared to those originally intended by the creators and their storytelling. Hence, there is a change not only in the reception but also in the transmission in the communication of auditory information. The player becomes a co-transmitter of information introducing non-linearities in the experience that propagate throughout the agents' chain of activity, triggering feedback and generating further non-linearities.

In this respect, Frauenberger's entanglement HCI (Sec. 1.3.2) suggests abandoning a user-centered design of the digital artifact in favor of participatory, speculative, and agonistic methods with the ultimate goal of obtaining meaningful relationships and not merely optimized processes relating to the human or the machine pole, or their interaction. It is useful to briefly recall these methods. The agonistic and **adversarial design** employs processes and creates spaces to foster vigorous but polite disputes involving designers' participation in order to constructively identify inspiring elements of friction [36]. On the other hand, the participation in a speculative process through designing **provotypes** aims to provoke a discussion about the technological and cultural future by considering creative, political, and controversial aspects [117].

The more degrees of freedom in the network configurations, the more behaviors can potentially be stimulated. The relational network should not be hardly controlled because its expressive potential can be exploited through its differentiation. In our opinion, the current immersive audio technologies are struggling to emerge, because they often introduce static agential cuts, justified by audio quality assessments conducted in a reductionistic way. On the contrary, the main goal of the digital twin is to favor the participation of all available configurations. Specific configurations and agential cuts emerge in a speculative, agonistic and provocative manner so that all actors can benefit from different attempts following **knowledge diffraction** [6]. The learning in such fluid and dynamic evolution from one configuration to another is a continuous flow of knowledge that informs the digital twin's activity. In other words, the digital twin continuously proposes new agential cuts to record and analyze the overall result. A relevant example in SIVE is the co-determination of the attentional focus in selecting the meaningful auditory information for a digital twin facing the cocktail-party effect [20]. The digital twin must be able to guide active participation with the VE considering listener's available knowledge extracted by previously experienced and stored scenarios (and agential cuts).

The continuous intra-action within the digital twin in relation to a shared and immersive experience is of strong practical relevance within the proposed theoretical framework. This issue offers concrete possibilities for radically changing the way we interact socially in the future, by using digital tools equipped with computational intelligence and **artificial intelligence** (AI) algorithms able to manage complex systems [107]. The decision-making phase of intelligent algorithms will improve over time, thanks to a dynamic identification and classification of configurations and links in the actor-network. The knowledge can be continuously extracted as a result of computational intra-actions of the *human-in-the-loop* type where the listener can be seen as an agent directly involved in the learning phase, step-by-step influencing cost functions and all other measures [69]. More in general, the reinforcement learning paradigm focuses on long-term goals, defining a formal framework for the interaction between a learning agent and its environment in terms of states, actions, and rewards, hence no explicit definition of desired behavior is required [35]. This process can be accomplished during exposure to a continuous stream of multimodal information like in the case of lifelong learning [109], or via interactive annotations and labelling [81].

## 1.4 A Taxonomy for SIVE

An important contribution to the design in VEs comes from practice, e.g. professional reports and testimonials, best practices, or reviews and interpretations of lessons learned in the industry (see Ch. 6 and [76]). Taking into account all these inputs, academic studies, new technologies, and commercial user feedback, different communities draw support for their specific users and domains of interest. Within the SIVE field, there is still much work to be done. There is a lack of recommendations and design analysis on building interfaces, interactions, and environments that fully exploit egocentric sonic information. To unlock such potential, our suggestion is to start from a multi and interdisciplinary work resulting in these foundational questions: does a development path exist for the SIVE field? Is an *ad hoc* theoretical approach necessary? Without going into the details of the epistemological crisis that is affecting the HCI field, we would try to avoid discussions on what is called in the HCI community *intermediate knowledge* [72] where positivist and constructivist perspectives are constantly clashing [45]. Examples of intermediate knowledge are all patterns/best practices proposed for certain aspects of the immersive experience.

Our theoretical framework uses an egocentric audio perspective by emphasizing the situated, embodied, enactive experience dimension of the listener with the different actors involved. An emphasis on the entanglement between humans and technology assumes that the listener's internal states are directly inaccessible to a non-intrusive and external technology, i.e. focused on exteroceptive sense [134].

There exist several classifications attempting to describe virtual spaces for sound and music purposes. The recent formulation in [4] distinguished three aspects:

- Immersive audio - the VE should provide the feeling of being surrounded by a world of sounds
- Interactive audio - the VE allows the user to influence the virtual world in some meaningful way
- Virtual audio - the virtual world must be dynamically simulated

They have already been extensively discussed in the previous sections and many of the existing taxonomies for VR [95, 134, 157] prioritizing the system (or simulation) or the user, not the close relationship with the listener. In this section, we propose an **audio-centered taxonomy** that does not distinguish between user and system, listener and simulation. We will motivate the selection of three dimensions able to describe a technological mediation in VR: **immersion**, **coherence**, and **entanglement**. The qualitative description in this section leaves as a future challenge a quantification of the performative processes introduced here.

Referring to the autobiographical element introduced in the book preface, the first meeting of the two chapter authors at the ACM CHItaly 2021, the biennial conference of the Italian HCI community, has also a scientific meaning. The paper by Geronazzo *et al.* [50] was presented 10 years ago, as one of the first tasks of the first author's doctoral program. He attempted to adapt the *virtuality continuum* of Milgram and Kishino [95] in the context of spatial audio personalization technologies for VR/AR. His main motivation was to overcome his difficulty in fitting the strong acoustic relationship (i.e. HRTF customization) between listener and technology into a taxonomy created for visual displays in 1994.

That paper proposes a characterization that uses a simplified two-dimensional parameter space defined in terms of the *degree of immersion* (DI) and *coordinate system deviation* (CSD) from the physical world. It is a simplification of Milgram's three-dimension space, summarized in the following:

- Extent of World Knowledge (EWK): knowledge held by the system about virtual and physical worlds
- Reproduction Fidelity (RF) - virtual object rendering: quality of the stimuli presented by the system, in terms of multimodal congruency with their real counterpart
- Extent of Presence Metaphor (EPM) - subject sensations: this dimension takes into account the observer's sense of presence

CSD matches EWK with the distinction that a low CSD means a high EWK: the system knows everything about the material world and can render the synthetic environment in a unified mixed world. From an ecological perspective, the system knows and dynamically fosters the overlap between real and virtual. On the other hand, EPM and RF are not entirely orthogonal and the definition of DI follows this idea: when a listener is surrounded by a real sound, all his/her body interacts with the acoustic waves propagating in the environment, i.e., a technology with high presence can monitor the whole listener's embodiment and actions (high DI).

Recently Skarbez *et al.* [134] have proposed a revised version of Milgram's virtuality continuum introducing two distinctive elements. First, a reduction from three to two dimensions similarly to [50]: Immersion and Extent of World Knowledge. In particular, Immersion is exactly based on the same idea as DI. Second, they introduced a discontinuity in the RF and EPM dimensions considering the absence of any display at the left side of the spectrum: the physical world without mediation is inherently different from the highest level of realism achievable through VR technologies that stimulate exteroceptive senses (i.e., sight, hearing, touch, smell, and taste). The latter consideration propagates to Immersion.

The rough taxonomy of Geronazzo *et. al.* missed the idea of coherence between simulation and human behavior, which is well identified as the third analytical dimension of Skarbez *et al.* [134]: *coherence*. It takes into account both plausibility and expectation of technological behaviors for the user in cognitive, social, and cultural terms. However, the three proposed dimensions cannot and do not claim to describe such a relationship between the user and the system as emphasized by the authors in their system-centered taxonomy. The work of Skarbez and colleagues is once again anchored to the distinction between user and system which generates several issues in framing the intra-actions of actors/factors in VR/AR sonic experiences.

To support the SIVE theoretical framework, we focus on purely VR only. This means that our discussion will not consider the CSD/EWK dimension assuming that there are no anchors to the physical world. However, since we are emphasizing the influence of human-real-world relationships on experience in VE and vice versa, we have decided not to make the world configurations explicit thus considering them as a whole with the listener. Extensions to mixed reality will be an object of future studies in a reviewed version of our theoretical framework.

Starting from the previously identified dimensions of Immersion [50] and Coherence [134], we suggest **three top-level categories** that need to be addressed through interdisciplinary design work. A schematic representation can be found in Fig. 1.5.

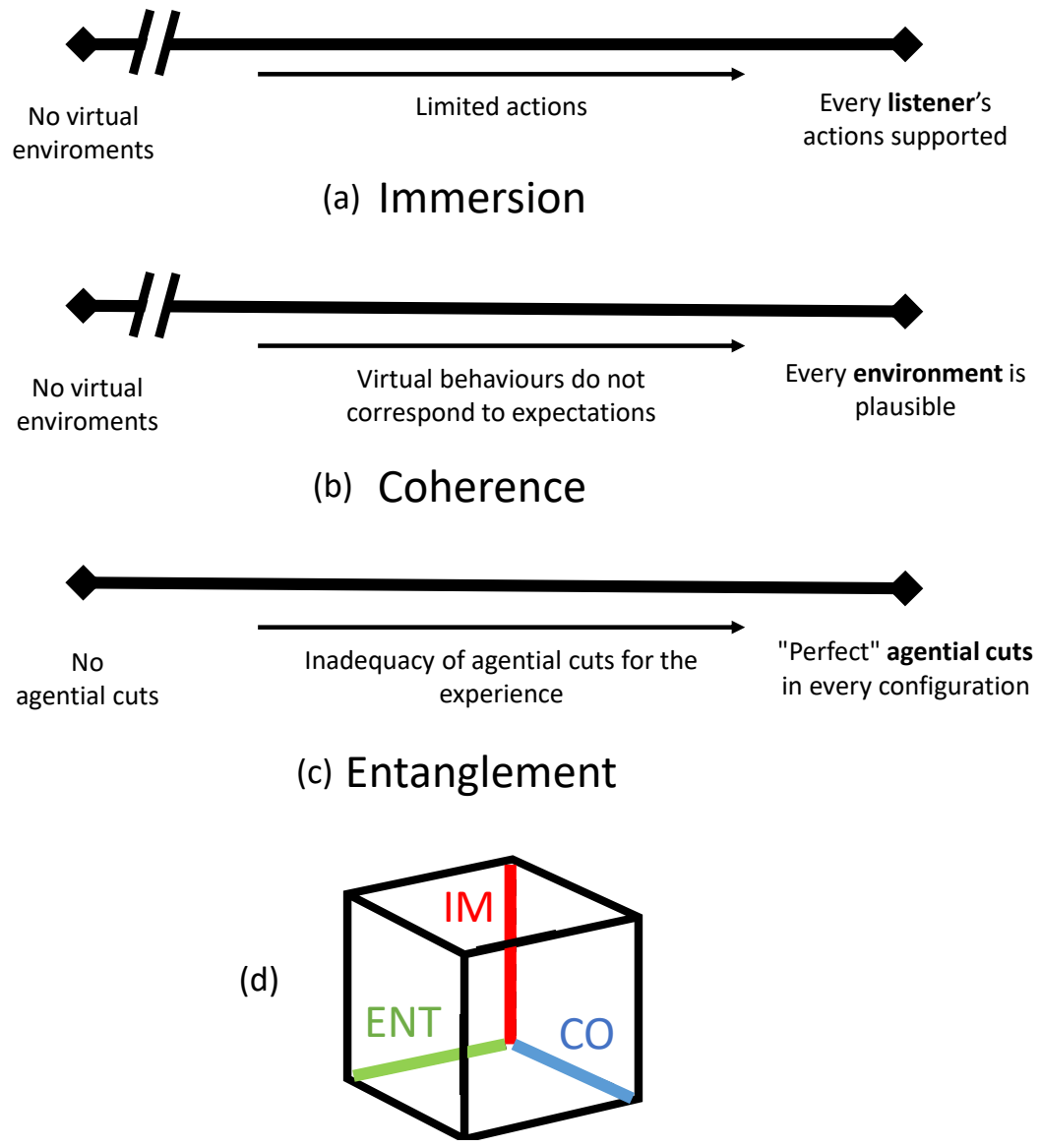

Fig. 1.5: Three-dimensional taxonomy for SIVE, (a) Immersion, (b) Coherence e (c) Entanglement, and their relations in (d).

**Immersion:** the digital information related to the listener-digital twin relationship supporting an increasing number of actions in VEs. It measures the technological level and its enactive potential between listener and auditory digital twin.

**Coherence:** the digital information related to the digital-twin-VE relationship that allows the plausible rendering of an increasing number of behaviors in VEs. It measures the effectiveness of sonic interaction design in VEs.

**Entanglement:** represents the overall effectiveness of the actor-network and its agential cuts that are dynamically, individually, and adaptively created. It measures participation in the locus of agency and its consequent phenomenological description. The auditory digital twin actively proposes new relations favoring redefinitions in the agential cuts related to the mutual transformative action between listener and technology.

To support our proposed taxonomy and framework for SIVE, we introduce a case study on a fictitious and purely theoretical artifact along the lines of *Flow* [45]. It allows us to decline the various facets in a flexible example.

*Spritz!* is an interactive and immersive VR simulation supported by full-body tracking, stereoscopic vision, and headphone auralization. It is designed to address the *cocktail-party problem*. The human selective attention requires different contributions and levels of perception in supporting the ability to segregate signals-also referred to as auditory signal analysis [15, 20]. When confronted with multiple simultaneous stimuli (speech or non-linguistic stimuli), it is necessary to segregate relevant auditory information from concurrent background sounds and to focus attention on the source of interest. This action is related to the principles of auditory scene analysis that require a stream of auditory information filtered and grouped into many perceptually distinct and coherent auditory objects. In multi-talker situations, auditory object formation and selection together with attentional allocation contribute to defining a model of cocktail-party listening [75, 132]. The design of *Spritz!* aims to give shape to an auditory digital twin able to detect listener intent, i.e., identify the relevance of a sound compared to other overlapping events. It can instantaneously determine the attentional balance within an auditory space. Its main goal is to promote the listener's well-being through manipulations of the sound scene in a participatory way respecting the listener's desires.

### 1.4.1 Immersion

According to Murray [98], the term ***immersion*** comes from the physical experience of being immersed in water. In a psychologically immersive experience, one aims at experiencing the feeling of being surrounded by a medium that is a reality other than the physical one, able to capture our attention and all our senses. Therefore, it has an important element of continuity with our framework by identifying a mediating action of VR experiences. According to Slater and Wilbur [137], the term **immersion is tightly linked to the technology**, the mediator, to elicit the sense of presence. Technological systems for immersive VR count several combinations of equipment and techniques, such as HMD, multimodal feedback, high frame rates, and large tracking areas. Such a heterogeneous arsenal is a complex system of functional elements that have an immediate impact on the listener's experience. Initially, technical specifications were reasonably identified as the main constraints for a VR experience. However, other elements were considered with a large-scale diffusion of VR technologies. The design of VEs became critical in those details that ensure a plurality of actions with virtual objects, the surrounding virtual world, and their representations. As discussed in [30], the effects of all these components are highly interconnected with each other. Moreover, the absence or misuse of any of them can produce immediate disruptions in the sense of presence or cybersickness [33], such as low headset quality [16] or unfiltered noise caused by sound sources external to the VR setup [136].

The strong connection between immersion and equipment means that different VR solutions hold an intrinsic level of immersion regardless of the actual applications performed with them [120]. This is evident when considering basic audio quality vs. quality of the listening experience. For instance, considering projected screens offers designers of VEs the opportunity to combine real and virtual elements in the tracked area (Chapter 13 offers an interesting reflection on artistic performances mediated by VR/AR technologies). However, the overall sense of presence experienced by the listener will depend on the specific combination of the HW/SW setup. Such setups support a certain type of action within the VE. The **Immersion "I"** dimension takes into account these features as the starting point of an enactive potential for the auditory digital twin. Such a potential intrinsically limits the development and creation of new actions.

Furthermore, the enactive egocentric perspective of Sec. 1.3.1 provides a solid theoretical framework for considering the importance of ecologically valid auditory information in eliciting a sense of presence in a VR-mediated experience. First of all, it should be mentioned that there is a lack of research related to the effects of interactive sound on the sense of body ownership and agency (refer to the discussion in Ch. 2). The vast majority of studies addressing presence from an auditory perspective focus on place illusion and spatial attributes. This should not come as a surprise, since many of these binaural attributes are perceived applying sensory-motor contingencies and embodied multisensory integrations. A simple example of spatial audio technologies is the importance of head movements data that are acquired by 3 degree-of-freedom head-trackers, allowing listeners to exploit binaural cues for resolving the so-called *front-back confusion* [22]. However, computational models for binaural cues are usually parameterized by the head radius or circumference, or ears position [52]. This example suggests that synchronization and plausible interactive variations, i.e., occurring in reaction to the digital twin's gestures in coherence with sensorimotor contingencies, can positively influence the sense of agency. In addition, other studies demonstrate how the **sound of action** and the active exploration can support haptic sensations and vice versa in a co-located and simultaneous manner. For instance, Ch. 12 analyzes the impact of sound in an audio-tactile identification of everyday materials from a bouncing ball.

Regarding spatial hearing, there is a huge differentiation in accuracy between more (experienced) and less (naive) reliable listeners [3, 51]. More generally, the distinction between categories of listeners is still challenging and is made based on several factors such as multisensory calibration and integration (see Ch. 12 for audio-haptics), familiarity with immersive/spatial audio technologies, musical background [152], or audio mixing experience, etc. (as in Ch. 9). Both acoustic (i.e. acoustic transformations of the body) and non-acoustic (i.e. everything else) factors are highly individual and depend on the relationship between the listener and the real world mediated by technology in a general sense (e.g. video games, musical instruments, etc.).

All **objectifiable information regarding the listener** is known configurations. For example, bottom-up approaches for modeling psychophysical phenomena of spatial hearing and multisensory integration fall into this category. Such knowledge has to be integrated into the immersive system, explicitly contributing to the actor-network managed by the digital twin.

Coming back to our *Spritz!* simulation, the level of **I** is expected to be high due to state-of-the-art technological components. The digital twin can recognize and manage several full-body skeletal configurations as well as near-field acoustics algorithms that take into account the acoustic coupling of the main joints such as the head and shoulders. This last aspect is usually largely underestimated in virtual acoustics systems [17]. The customization based on anthropometry allows the digital twin to guide the acoustic rendering of movements considering head tilt and torso shadowing in real-time. Furthermore, binaural and spectral cues might be personalized and weighted according to the listener's level of uncertainty, allowing the digital twin to predict which sound sources are most likely to be segregated based on an egocentric direction-of-arrival perspective.

The contribution of the **I** dimension can be summarized as follow: **I** is the digital information related to the listener-digital twin relationship limited by a specific technological setup. The support of an increasing number of actions in VEs is a consequence of technological improvements (both HW/SW) and/or an increasing objectification of the listener's configurations. Considering the idea of immersive potential of Ch. 11, limitations in enaction determine which changes are significant after technological manipulation. The level of reconfigurability within the digital twin accounts for the constant dialogue with the listener to explore her state and tendency to immersion in every moment of the experience (see also Sec. 1.4.3).

### 1.4.2 Coherence

The VR simulation must be able to make the digital twin freely interact with the VE, eliciting a plausible experience for the listener who is always aware of the mediated nature of the experience. In other words, the interaction design must support functionally and plausible actions, the 'doing' [43]. This means that possible configurations of the technical setup and the listener (the objectification in the digital-twin, see Sec. 1.4.1) constitute the enactive potential of immersion and must be balanced within the sonic interactions.

In this section, we focus on the ***coherence*** of the digital-twin/environment relationship. On the other hand, Sec. 1.4.3 provides an interpretation of the dialogue with the ***immersion*** dimension.

VE simulations can create fictional worlds, exploiting opportunities for both naturalistic and magical interactions [13]. Designers can experiment with defining rules that only apply in the virtual domain, such as scale, perspective, and time. The philosophical discussion of the dualism "Doing vs. Being" in [4] provides interesting insights for our egocentric auditory perspective: simulation can have different levels of interactivity suggesting different action spaces for the digital twin in the virtual worlds.

Interacting with the VE, avatar included, consists of **altering the states of 3D elements** that have been created at different levels of proximity: the virtual body (i.e. avatar), the foreground (i.e. peripersonal object manipulation space), and in the background (i.e. extra-personal virtual world space). Existing researches on 3D interaction focuses on the spatial aspects of the following main categories: selection, manipulation, navigation, and application control (the latter involving menus and other VE configuration widgets). Selection techniques allow users to indicate an object or a group of objects. According to the classification of Bowman *et al.* [14], one can consider selection techniques based on object indication (occlusion, object touch, pointing, indirect selection), activation method (event, gesture, voice command), and feedback type (text, acoustic, visual, force/tactile). Manipulation techniques allow the digital twin to modify all virtual objects configurations that are made accessible to it: e.g. the spatial transformation of objects, i.e. roto-translation and scaling, surface properties such as material texture and acoustic properties, or 3D shape and structure manipulations. For the disparate interaction metaphors for selection, we refer to a recent review in [92]. Finally, navigation techniques allow digital twins to move within the VE to explore areas and virtual worlds. Typical movements include walking and virtual transportation, including flight experiences. In particular, walking is fundamental to humans, and supporting natural locomotion is not always feasible on a limited tracked space. Accordingly, there are other interaction metaphors such as walk-in-place [42], teleportation, or semi-automatic movements between control points [61]. It is worthwhile to mention the self-motion illusions. In circular vection [116], moving sounds surrounding the listeners facilitate the perception of being in motion when in fact they are not. For spatial design considerations in sonic interactions, Ch. 6 provides a comprehensive analysis and a typology of VR interactive audio systems.

These configurations must be plausible and the digital twin should support a dynamic transition from one to another. This is crucial to avoid irreparable breaks in presence. Therefore, **coherence "C"** describes the degrees of freedom introduced by the sonic interaction design in VEs based on the active dialogue between the digital-twin and the VE, established experience after experience.

In this section, we are particularly interested in the plausibility illusion determined by the overall credibility of a VE concerning subjective expectations. It is not only a coherence between external events not directly caused by listeners but an objective feature of the VE [134]. Its reconfigurability includes an **internal logical coherence and a behavioral consistency** considering prior knowledge. Sound conveys ecological information relevant to the expectation towards VE behaviors compared to the listener's everyday experience: embodied, situated in a socio-cultural context. The environment configurations (avatars and virtual worlds) intertwine with the known listener configurations held in the digital twin. Once again, the digital twin has a central role in acting following an egocentric audio perspective (see Fig. 1.4 for this foundational idea). Dimension **C** advocates a top-down approach to interactions, constituted of cognitive and sociocultural influences based on listener real life.

Moreover, coherence does not presuppose physical realism. It fosters interactions in coherent virtual magic worlds. The dynamic dialogue between VE and digital-twin makes it possible. For example, let's consider a cartoon world where simplified descriptions of sound phenomena exaggerate certain features [118]. It may be plausible as long as it conveys relevant ecological information. Audio procedural models are based on simplifications in properties and behavior of a corresponding real object, i.e. simplified configurations. Such parameterization can be informed by auditory perception and cognition maintaining ecological validity of a fictional sonic world while reinforcing the listener's sense of agency. Digital information regarding the relationship between the digital twin and VE allows the creation of an increasing number of plausible behaviors in VR.

Considering once again the distinction among avatar, peri- and extra- personal spaces, neurophysiological research on body-ownership and multisensory integration suggests the existence of a fluid boundary in the **perceived space by subjects** [60]. It is worth noticing that the neuronal activity sensitive to the appearance of stimuli within the personal space is multisensory in nature and involves neurons located in the frontoparietal area. In this area, neuronal activity is related to action preplanning particularly for reacting to potential threats [130] and elicit defensive movements when stimulated [32]; these multimodal neurons combine somatosensory with body position information [58]. Bufacchi and Iannetti [24] suggested that the personal space should be described as a series of action fields that spatially and dynamically define possible responses and create contact-prediction functions with objects. Such fields may vary in location and size, depending on the body interaction within the environment and its actual and predicted location. Space is also modulated in response to external stimuli and internal states of the subject, defining a relationship between listener, environment, and tools [119].

Of particular interest are modulations due to the **proxemics**. The term was introduced by Hall [63] and concerns implicit social rules of interpersonal distance among people conveying different social meanings. The cooperation in a socially shared interpersonal space [144] requires to support the transition from individual to collaborative spaces [142]. In Ch. 8, the design of sound intensity (or sound attenuation) as a function of the proximity from a sound source is addressed. Different configurations of personal and public spaces were tested in a shared VE for collaborative music composition. Interestingly, rigid boundaries in the transition between spaces forced listeners to take a **social distance** and isolate with a negative impact on the collaborative aspect moments of the composition process. Therefore, the separation between public and personal space should be fluid rather than rigid. The VE should be configurable in the social aspects that emerge from the strong interconnection between configurations made available to the digital twin, increasing the fluidity and better supporting collaboration in shared experiences.

In *Spritz!* scenario, we should identify the VE's abilities in shaping the simulation within the digital twin. First, *Spritz!* Has multiple configurations accounting for different strategies of the level of audio details. The radial distance with an egocentric reference can drive the dynamic definition of three partially overlapping levels of detail associated with proximity profiles: avatar, personal and public. The avatar's movement sounds are rendered through procedural approaches with individualized configurations based on listener acoustics; the personal space can manipulate sound behavior with simplified models taking into account security and privacy levels required by the situated and embodied states of the digital twin. Finally, sounds in the public space can be clustered, grouped, or attenuated by implementing plausible statistical behavior, e.g. using audio impostor replacement such as audio samples.

The *Spritz!* environment should facilitate resolutions of the cocktail-party problem in crowded situations. Accordingly, it should be able to apply noise suppression of negligible information in the public space or vice versa to operate audio enhancements supporting attentive focus. This dynamic connection between VE and digital twin should be able to maintain *coherence* in the induced behaviors, supporting actions plausibility while bending the space around the listener.

A meaningful manipulation of virtual spaces is crucial and creative. Since SIVE naturally includes music composition, a VE must foster the development of individual or collaborative creative ideas through dynamic control of its configurations within and by the digital twin. In particular, results in Ch. 8 support VE spatial design as the creation of "magical" exploratory opportunities, adding original dynamics to collaborative work in VEs. The digital twin has a pivotal role in such space modulation that allows tracing boundaries performatively and eliciting internal emotional states following the listener/composer's expectations.

### 1.4.3 Entanglement

The listener's susceptibility to immersive VE experiences is usually determined by administering questionnaires [155, 156]. The experimenters' aim is usually to perform a screening test to distinguish who can and will be able to easily immerse in a VR mediated situation. Furthermore, this separation is assumed to remain constant throughout a short-enough experiment. The immersive tendency can change over time due to training, learning, experience, mood changes and personality, etc. (see Ch. 11 for further details). For such reasons, common recommendations in conducting VR experiments suggest conducting single experimental sessions. However, studying the impact of the aforementioned dynamic changes opens to the third and last dimension of our taxonomy: ***entanglement***, which is the knowledge extraction from the evolution of an actor network able to reveal multiple facets of the egocentric experience in time, space, and intra-action.

The first step requires describing the available configurations. Starting from the idea of immersive tendency, VR simulation would benefit from the **knowledge of the listener's susceptibility** towards configurations of setup and environment to modify or avoid non-significant experiences, e.g. getting a break in illusion. In other words, (quantifiable) listener configurations must be defined, discovered, and actively explored by the digital twin. For example, the way sound samples are engineered is very interesting here. A sliding friction sample, e.g., squeaking, rubbing, etc., requires a large amount of data and randomization techniques to avoid repetition. Sounds should be consistent with the listener's expectations in response to complex and continuous motor actions. For this reason, procedural audio approaches can tightly connect the sound to complex and continuous motor actions.

The **Entanglement** dimension ("**E**" ) aims to provide a phenomenological characterization of actors' evolution and activities based on their performativity and participation in a locus of agency. We realize the high complexity of such a descriptive and formal process, but we believe that an attempt in capturing the **transformative potential of VR** mediated experiences is worthwhile to be conducted for the SIVE discipline. Of great importance here is the idea of **monad** by sociologists Tarde and Latour [84, 143]: *"A monad is not a part of a whole, but a point of view on all the entities taken severally and not as a totality."* One can consider a monad as a relational perspective of each actor, shifting the emphasis from aggregation of the whole to movement between different points of view. The main purpose of any perspective is the structural analysis of the network and its configurations and, at a later stage, to derive knowledge and understanding of its dynamics. The inherently egocentric local perspective of the locus of agency, i.e. digital twin, is again emphasized as opposed to a global view. Egocentric networks built around specific nodes such as the listener configurations can support the exploration of intra-activated dynamics. Configurations and links can be discovered and/or modified during different mediated experiences.

The **actors' collaboration is vital in integrating different points of view**, creating opportunities for meaningful experiences. In shared VEs (Ch. 8), listeners are co-present with other human participants interacting in an interpersonal way. Research interests of computing-supported cooperative work can provide interesting insights into prioritizing collaboration [99]. The choice of collaborative models fostering the design of active VEs for meaningful and creative experiences is of particular relevance to entangled SIVE.

Intentionality and gesture support can be achieved through continuous network reconfiguration. Identifying common goals through inter-actor communication are fundamental requirements to increase the digital-twin enactive potential. We argue that this area of research is new for SIVE, especially in these collaborative aspects. Many fundamental and critical questions for SIVE are waiting to be answered.

Digital transformation promotes ubiquitous and pervasive interconnected data sets with the opportunity to offer new ways of navigating and extracting knowledge. Dork *et al.* [37] explored the visualization of relational information spaces, incorporating both the individual and the whole in a monadic perspective. The authors' goal was to exploit the rich semantic connections to design new exploration methods for inter-connected elements. There is an increasing interest in more exploratory forms of information retrieval without specific needs/constraints, sustained by the **desire to learn, play and discover** openly [91]. In analogy with these practices, the digital twin should curiously move between nodes, configurations, and connections experimenting and manipulating the actor-network for sense-making. To encourage surprising discoveries and interest within experiences, the digital twin should offer unconventional and appealing views with agency.

The auditory digital twin actively proposes new relationships and encourages agential cuts definitions under the mutual transformative action between listener and technology. In the monadic perspective of the digital twin, the distinctive qualities of each actor within a VE should be situated emerge. Differentiation among configurations is not an a priori actor property but its unique position in the network. Each actor imprints its particular identity on an ever-changing relational world. In other words, the digital twin is looking for differences in each actor by considering different monadic perspectives. VR simulations allow us to take the point of view of the elements thanks to shared virtual world knowledge.

In the area of AI agents, i.e. non-human entities capable of interacting with ecological behaviors [109], intelligent algorithms would have the predictive potential on the listener's action programme. Their ability in monitoring and predicting listeners' behavioral responses could enable the digital twin to determine listeners' expectations and cognitive and psychological capabilities [25]. Moreover, AI algorithms could propose exploration paths to the listener within VEs. Therefore, the capabilities of safely navigating through temporary, transient, and overlapping configurations are definitely complimentary to predictive power.

In line with the emerging research area called **immersive analytics,** humans and AI can support each other in decision-making based on the navigation in shared thinking spaces [133]. Meetings between the listener and her digital twin can take place in a virtual meta-environment where configurations and connections of an experience can be a posteriori analyzed, collaboratively. The unique personal supervision of the AI algorithms implemented in the digital twin could reflect the listener's traits and interests. Understanding the listener's preferences and assessing their impact on the predictive performance of AI algorithms can help to propose adaptive and customizable systems with a certain level of memory of past VR mediated experiences [103].

Finally, how can we measure the overall effectiveness of an actor-network and its agential cuts that the digital twin dynamically, individually, and adaptively creates? This question corresponds to Latour *et al.*'s challenge to take into account long-term features, indicative of an order that might be learned navigating overlapping perspectives (monads) [84]. Such an emphasis on navigation gives a unique role to **movement/exploration** as a way of experiencing relationships and differences between configurations. Therefore, we suggest that the digital twin should navigate along with different and novel perspectives for sense-making. The dynamic relational quality of each actor's unique position in network space, i.e. agential cuts, reflects the exploration potential shaping and creating meaning for the listener.

We argue that the VR mediated experience is never solitary, considering both human and non-human actors. Any actor cooperates within a shared VE, e.g. to perform a musical performance (Ch. 13) or a spatialized audio mixing (Ch. 9). Collaboration takes place on a common task, which has a huge impact on the intra-action dynamics. In addition to the exploratory movements, technological transparency introduced in Sec. 1.3 is a key factor influencing "E" measures. In analogy with the sense of presence, **co-presence** [26], i.e. the feeling of sharing a VE with others, has been shown to strongly depend on avatar appearance and its realism, as well as on the cooperation level in task completion [111]. A second aspect recalls **awareness** [7] which is the action understanding of other actors, especially in non-human agents. This latter concept strongly relates to trustworthy AI issues and explainable AI [70].

A further "E" measure in SIVE can be inspired by the River and MacTavish's framework [117]. They proposed to generate low-level prototypes of an artifact from simplified attributes. The more extreme the change in such attributes, the more likely the change will be to provoke and reveal hidden assumptions in the design process. In our taxonomy, we call it generative potential in explorative movements and network changes, and technological transparency.

The final example in our fictitious case study *Spritz!* considers the meaningful prediction of the listener's intentionality and understanding of other sources of interest, e.g. avatar's gestures or other avatars. *Spritz!* should be able to support attentional focus. A virtual ray/cone pointer projected by the avatar through the VE or a virtual cursor/hand mapped to the listener's body movements might facilitate the selection of points of interest. Gesture analysis could provide *Spritz!* relevant information for semi-automatic focus support. This scenario opens to the experimentation and development of " magic" interactions of virtual superhuman hearing tools such as dual-beamformers guided by the avatar's body [53]. *Spritz!* should be free to propose novel ways of interaction and exploration within VEs. This dynamic dialogue can be considered a form of **virtual provotyping** that has to guarantee *coherence* with all available sensorimotor contingencies, having a positive effect on the listener's sense of agency in any proposed behaviors.

## 1.5 Conclusion

This chapter aims at emphasizing how the SIVE book was born and developed in a constantly evolving situation in the field of human-computer interaction. We invite the reader to explore all its chapters with this shared and dynamic tension that we, as editors, have tried to formalize in what we have called the egocentric perspective of the auditory digital twin. The co-transformation of man and technology seems to us a central theme that will surely help us to enter the $4^{th}$ HCI wave, consciously.

The proposed taxonomy focuses on action, behavior, and sense-making because we believe it is a meaningful way for authentic auditory experiences in VR. In particular, the last aspect of sense-making turns out to be the most challenging. The idea of diffraction and exploration of differences and discoveries requires novel ways of scientific investigation in SIVE. The most crucial might be the level of personalization that future technologies will require from the listeners. New paradigms for artificial and immersive interaction between humans and VE will have to be proposed. The attribution of agency to a digital twin is a network effect that will have relevant ethical implications, as well as complexity in its analysis.

How much would I trust my digital twin? Its intermediary role, sometimes provocative, in search of differences can elicit strong reactions in the listeners. Will the listener accept and share this perspective? The affective information strongly links sound to meaning [138], creating empathy between listener and digital twin. This aspect will be carefully considered for its ethical implications.

How can one quantify and classify the various actor networks in the proposed three dimensions? Surely, this is an open challenge of this first proposed theoretical framework for SIVE. Visualizing and representing transitions and agential-cuts are relevant issues toward an objective description of mediation phenomena. Creating multiple ontologies in " magical" interaction metaphors allows to transcend reality and immerse into unique experiences within VEs. Since VR is not yet able to fully replicate natural reality and may not be able to do so, its current features actually allow listeners to do and be things that are impossible in the real world. This is the very essence of knowledge diffraction: the digital twin should explore such differences that are impossible to test in the physical world, extracting meaning for the listener. Of particular interest here, the ideas of superhuman powers and virtual prototyping [53] reflect human desire to increase our capabilities. They are receiving increasing attention thanks to the post-humanism and human enhancement manifestos [97]. Following this line of thought, Sadeghian *et al.* [121] proposed to VR designers to explore new forms of interaction without necessarily imitating the physical world. VR's limitations in creating realistic interactions are replaced by a focus on experiences that are impossible to have in the real world, such as superhuman powers of flying, X-ray vision, shape-shifting, super memory, etc. . Limitations obviously occurred while differentiating VEs before confusion invades the listener. Indeed, a balance in ecological and familiar stimulation should guide the creation of a "safety net" for the listener – the digital twin's exploration of agonistic and provocative knowledge opportunities without shortcoming.

This chapter aims to shape the SIVE research field, **sonic interactions in VEs**, that is ready to welcome wide-ranging reflections on what might be called

**sonic intra-actions in VEs**.